\documentclass[aps,prc,showpacs,nobibnotes,final]{revtex4}
\usepackage{graphicx}
\begin{document}

\title{Reduction of superfluid gap by scattering}
\baselineskip=1. \baselineskip

\author{     P. Bo\.{z}ek}
\email{bozek@solaris.ifj.edu.pl}  
\affiliation{ National Superconducting Cyclotron Laboratory, and Department 
of Physics and Astronomy, Michigan State University, 
East Lansing, MI-48824 \\ and \\
 Institute of Nuclear Physics, PL-31-342 Krak\'{o}w, Poland}
\date{\today}



\begin{abstract}
The effect nucleon dressing   by scattering
on the pairing gap in nuclear matter is discussed. Numerical results
from self-consistent T-matrix calculations are compared to
quasi-particle approximations. The dominant effect of scattering can be 
accounted 
for by a renormalization  of the quasi-particle strength. 
The
 exponential sensitivity of the pairing gap and the critical temperature on 
the pairing strength makes it strongly dependent on many-body effects.
\end{abstract}



\maketitle




\section{Introduction}

Nuclear systems at low temperatures undergo a superfluid phase transition.
This is observed in finite nuclei as the even-odd staggering of masses and
is expected to occur in infinite nuclear matter inside neutron stars.
Calculations in finite nuclei usually assume an effective pairing 
interaction fitted to the available data. Calculations in nuclear matter
are using predominately the bare nucleon-nucleon interaction in the
gap equation. 

Due to the short range repulsive core in the nuclear potentials 
Brueckner type ladder resummation of the  interaction in medium 
is necessary.
On the other hand, the effective interaction in the gap equation should
be two-particle irreducible, i.e. without resummation of the
 particle-particle  (and hole-hole) ladder \cite{migdal}. Thus a good starting
point could be the bare nucleon-nucleon interaction, the same as used in the
ladder approximation for the two-particle correlation.

The value of the pairing gap depends on nucleon single particle energies. 
Using single-particle energies obtained from Hartree-Fock or 
Brueckner-Hartree-Fock (BHF) 
calculation a different value of the paring gap is 
obtained.
It can be understood as a modification of the effective mass,
 and consequently a  change in  the density of states at the Fermi energy.
Obviously the value the effective mass is an inherent part of effective
parameterizations of the pairing interactions used in calculations
 in finite systems.

Only relatively few works discuss  possible in medium modifications of the
pairing interactions beyond a change in the effective mass.
Polarization
 corrections to the bare nucleon-nucleon interactions were discussed
in Refs. \cite{wambach}. It was found that the screening reduces
 the pairing gap by a factor $\simeq 3$. Another in medium modification was 
analyzed in the framework of  self-consistent nuclear matter calculations 
\cite{ja2}. Self-consistent nuclear matter calculations use off-shell
 nucleon propagators in the  resummation of ladder diagrams for
the self-energy. In this way a self-consistent spectral function can be
 obtained.
It was found that the use of full spectral functions in the 
 gap equation leads to a strong reduction of the critical temperature 
and of the superfluid gap in comparison to the quasi-particle approximation.
It is the goal of the present paper to identify the main cause of  this
modification of superfluid properties of  nuclear matter and to propose 
 a renormalization of  the quasi-particle approximation for the gap equation.
In Sect. \ref{normalsec} we discuss modifications of the two-particle 
correlations in the normal phase. This enables us to calculate the critical
 temperature, at which long range two-particle pairing
 correlations appear. The critical temperature is estimated using  
full
 spectral functions and in the renormalized  quasi-particle approximation.
In Sect. \ref{supersec} we present a discussion of the gap equation with 
full spectral function and compare it to its quasi-particle limit and to
 the usual mean-field gap equation.
In the concluding section \ref{conc}
  we identify the most important renormalization
 of the pairing interaction due to off-shell propagation and indicate 
its consequences for realistic nuclear matter calculations.

\section{Quasi-particle limit of self-consistent ladder resummation}

\label{normalsec}

Nuclear medium is a relatively dense system of particles strongly 
interacting on short distances.  Brueckner 
resummation of particle-particle ladder 
diagrams defines the so called in medium G-matrix 
\begin{eqnarray}
\label{gmatrix}
<{\bf p}|G
({\bf P},\Omega)|{\bf p}^{'}>& =& V
({\bf p},{\bf p}^{'})
 + 
\int\frac{d^3q}{(2 \pi)^3} V
({\bf p},{\bf q}) 
\frac{(1-f(\omega_{p_1}))(1-f(\omega_{p_2}))}
{\Omega-\omega_{p_1}-\omega_{p_2}}
 <{\bf q}|G({\bf P},\Omega)
|{\bf p}^{'}> \ ,
\end{eqnarray}
where ${\bf p_{1,2}}={\bf P}/2\pm {\bf q}$.
G-matrix resummation  allows to define 
single particle energies and gives  relatively good results
for the saturation properties of nuclear matter.
In the above equation and in the following we skip the spin, isospin indices
which are implicitly summed over.
The above equation corresponds to a resummation of particle-particle 
ladders, with medium effects entering through the Pauli blocking factors 
$1-f(\omega_p)$ in the numerator and single-particle energies $\omega_{p}$ 
in the denominator. Most advanced calculations in the Brueckner scheme use 
the so called continuous choice for the single particle energies $\omega_{p}$,
self-consistently defined by the G-matrix \cite{bhf}.

Another approach starts from the T-matrix approximation for the two-particle 
correlations \cite{vo,roepke}. In this scheme the ladder diagrams include 
both particle particle and hole-hole propagation. The Pauli blocking factor 
$(1-f(\omega_{p_1}))(1-f(\omega_{p_2}))$ in the G-matrix equation 
is replaced 
by  $1-f(\omega_{p_1})-f(\omega_{p_2})$ in the equation for
the retarded T-matrix
\begin{eqnarray}
\label{tmatrixqp}
<{\bf p}|T
({\bf P},\Omega)|{\bf p}^{'}>& =& V
({\bf p},{\bf p}^{'})  + 
\int\frac{d^3q}{(2 \pi)^3} V
({\bf p},{\bf k}) 
\frac{(1-f(\omega_{p_1})-f(\omega_{p_2}))}
{\Omega-\omega_{p_1}-\omega_{p_2}+i\epsilon}
 <{\bf q}|T({\bf P},\Omega)
|{\bf p}^{'}> \ .
\end{eqnarray}
The  imaginary part of the retarded 
self-energy in the T-matrix approximation is 
\begin{eqnarray}
Im \Sigma(p,\omega)=\int \frac{d^3k}{(2 \pi)^3} 
<({\bf p-k})/2| Im T(|{\bf p+k}|,\omega_k+\omega)|({\bf p-k})/2>_A
\Big(f(\omega_k)+b(\omega+\omega_k)\Big) \ ,
\end{eqnarray}
where $b(\omega)$ is the Bose distribution. $<\dots>_A$
 denotes antisymmetrization 
of the T-matrix (also in the spin, isospin indices not explicitly show).
The real part of the self energy  consists of the Hartree Fock self-energy 
and a dispersive contribution obtained from $Im\Sigma$
\begin{equation}
\label{disper}
Re\Sigma(p,\omega)=\Sigma_{HF}(p)+{\cal P}\int\frac{d\omega^{'}}{\pi}
\frac{Im\Sigma(p,\omega^{'})}{\omega^{'}-\omega} \ .
\end{equation}
The imaginary part of the self-energy is usually neglected leading to the 
quasi-particle approximation for the two-nucleon propagator in the 
T-matrix (Eq. \ref{tmatrixqp}).

Allowing for off-shell propagation of  nucleons and taking the self-energy
self-consistently (also its imaginary part)
 requires the use of full spectral functions in the 
calculation resulting in a  more complicated expressions for the T-matrix and
 the self-energy \cite{d1,ja}
\begin{eqnarray}
\label{teq}
<{\bf p}|T({\bf P},\Omega)|{\bf p}^{'}>& =& V({\bf p},{\bf p}^{'})
\nonumber \\ & & + 
 \int\frac{d\omega_1}{2\pi}\int\frac{d\omega_2}{2\pi}
\int\frac{d^3q}{(2 \pi)^3} V({\bf p},{\bf q})
\frac{\big(1-f(\omega_1)-f(\omega_2)\big)}
{\Omega-\omega_1-\omega_2+i\epsilon}
A(p_1,\omega_{1})A(p_2,\omega_{2})
 <{\bf q}|T({\bf P},\Omega)
|{\bf p}^{'}> 
\end{eqnarray}
and
\begin{eqnarray}
\label{ims}
{\rm Im}
\Sigma^+(p,\omega)& =&\int\frac{d\omega_1}{2 \pi}\int \frac{d^3k}{(2 \pi)^3}
A(k,\omega_1) <({\bf p}-{\bf k})/2|{\rm Im}T({\bf p}
+{\bf k},\omega+\omega_1)|({\bf p}-{\bf k})/2>_A
 \Big( f(\omega_1)+g(\omega+\omega_1) \Big) \ .
\end{eqnarray}
Equations (\ref{teq}), (\ref{ims}) and (\ref{disper}) have
 to be solved iteratively with the constraint on the assumed density $\rho$
\cite{ja,ja2}
\begin{equation}
\rho=\int\frac{d\omega}{2\pi}\int\frac{d^3p}{(2\pi)^3}
A(p,\omega)f(\omega) \ .
 \end{equation}

Results of the self-consistent calculation and of the quasi-particle
approximation are very different. In medium cross sections for nucleon-nucleon
 scattering are smaller when using off-shell nucleons \cite{d1}.
Also the critical temperature is strongly reduced when using full spectral 
functions \cite{ja2}.

The T-matrix approximation for the two-particle propagator is directly
related to the superfluid gap properties by the Thouless criterion 
\cite{thouless} and the condition for long range order \cite{km,hau2}.
On the other
BHF calculation using the G-matrix are are much developed for realistic 
interactions. In this work we derive an improved gap equation which could use
the results of advanced G-matrix calculations as an input.

\subsection{Thouless criterion for superfluidity}

\label{Thouless}
 The critical temperature can be obtained from 
the Thouless criterion for superconductivity \cite{thouless}. 
This is the temperature where 
a singularity in the T-matrix appears at twice the Fermi energy ($\Omega=0$)
 and zero
total momentum of the pair ($P=0$).
It means that the real part of the inverse T-matrix develops a 
zero-eigenvalue at the critical temperature
\begin{equation}
\label{intgap}
\int \frac{ d^3 p^{'}}{(2 \pi)^3}
<{\bf p}|{\rm Re} T
^{ -1}({\bf P=0},\omega=0)|{\bf p}^{'}> \Delta
({\bf p^{'}}) = 0 \ .
\end{equation}
It is equivalent to the existence of a nontrivial solution of the
gap equation at $T_c$
\begin{eqnarray}
\label{gapfull}
& & \Delta
({\bf p}) =  \nonumber \\ & & -
 \int\frac{d^3k}{(2 \pi)^3} \int \frac{d \omega}{2 \pi}
\int \frac{d \omega^{'}}{2 \pi} 
 V({\bf p},{\bf k}) 
\frac{A({ k},\omega-\omega^{'})
A({ k}, \omega)\big(1-f(\omega-\omega^{'})-
f(\omega^{'})\big)}{\omega} 
 \Delta({\bf k}) \ . 
\end{eqnarray}
The two  propagators in the gap equation above enter with the full spectral
function.
 On the other hand the BCS quasi-particle gap 
equation is
\begin{eqnarray}
\label{gapzwykk}
 \Delta
({\bf p}) + 
 \int\frac{d^3k}{(2 \pi)^3} \
 V({\bf p},{\bf k}) 
\frac{\Big(1-2f(\zeta_k)\Big)}{2\zeta_k} 
 \Delta({\bf k}) = 0 \ 
\end{eqnarray}
with $\zeta_p=p^2/2m+\Sigma_{HF}(p)-\mu$, and it 
corresponds to the Thouless criterion for the quasi-particle T-matrix,
i.e. the appearance of a singularity in the T-matrix given by Eq.
(\ref{tmatrixqp}). 
Because of additional averaging over spectral functions in the
gap equation with off-shell propagators, a different 
value of critical temperature comes out.

 The self-consistent T-matrix calculation was done for 
a simple, S-wave, interaction  and compared to the quasi-particle 
approximation for the gap equation \cite{ja2}.
 At a density of $0.45$ of normal 
nuclear density $\rho_0$
 it was found that the critical temperature was reduced from
$T_c=5$MeV in the mean-field gap equation with Hartree-Fock 
single particle energies to $T_c=1.6$MeV in the T-matrix approximation with
off-shell propagation \cite{ja2}. 
Below $T_c$ a modified T-matrix resummation was used in Ref. \cite{ja2}.
The resulting superfluid gap is significantly smaller than the
one obtained from the usual gap equation with quasiparticles. Modifications
 to the gap equation below $T_c$ coming from the use of 
 full spectral functions will be discussed
in Sect. \ref{supersec}. Here we concentrate on the T-matrix equations in the
normal phase, which is sufficient for the calculation of the
critical temperature. The simple Yamaguchi S wave interaction \cite{ja2} 
does not permit calculations at low temperature for normal density nuclear 
matter.

\subsection{Renormalized quasi-particle interactions}

Excitations in the Fermi liquid close to the Fermi energy can be described by 
quasi-particles. Quasi-particles are
 propagating on shell with dispersion relation 
modified by the presence of the medium. Also the scattering amplitudes 
between quasi-particles are modified by the medium. One of these modification
comes from the quasi-particle limit in the propagator of two nucleons
\cite{migdal,pn}.
Two particle Green's function can be formally written as a resummation
of particle-particle (and hole-hole) ladder diagrams starting
from in medium two-particle irreducible vertex \cite{migdal,abrikosov}.
For short range interactions the two-particle irreducible vertex can be 
approximated in the lowest order by the bare nucleon-nucleon interaction,
leading to the T-matrix equation (\ref{teq}). It should be pointed out that 
due to the instantaneous form of the interaction in the T-matrix equation,
the full T-matrix depends only on the total energy, and its equation 
takes a simple form for the  retarded T-matrix (Eq. \ref{teq}), 
also at finite temperature.

Close to the Fermi energy the spectral function becomes peaked around the
quasi-particle pole
\begin{equation}
A(p,\omega)=Z_p 2 \pi \delta(\omega-\omega_p) + R(p,\omega) \ ,
\end{equation}
where
\begin{equation}
\label{zfac}
Z_p=\Big(1-\frac{\partial Re\Sigma(p,\omega)}
{\partial \omega}|_{\omega=\omega_p}\Big)^{-1}
\end{equation}
and
$R(p,\omega)$ is the regular part, smooth in the vicinity of the 
quasi-particle pole.
The retarded propagator of two nucleons appearing in the T-matrix ladder
can be written as
\begin{eqnarray}
 \frac{\big(1-f(\omega_1)-f(\omega_2)\big)}
{\Omega-\omega_1-\omega_2+i\epsilon} 
A(p_1,\omega_1)A(p_2,\omega_2)& =&  B^{reg}(p_1,\omega_1,p_2,\omega_2,\Omega)
\nonumber \\& &
+ (2 \pi)^2 \delta(\omega_1-\omega_{p_1})\delta(\omega_2-\omega_{p_2})
 \frac{Z_{p_1}Z_{p_2}\big(1-f(\omega_{p_1})-f(\omega_{p_2})\big)}
{\Omega -\omega_{p_1}-\omega_{p_2}}  \ .
\end{eqnarray}
 $B^{reg}$ describes the part of the propagator of two nucleons
which cannot be written as propagation of two quasi-particles, it 
originates from the background part $R(p,\omega)$ of the spectral function.
Besides this contribution, the propagator of two-nucleons differs from the one
 used in the quasi-particle T-matrix (\ref{tmatrixqp}) by the presence of 
 renormalization 
factors $Z_p$ (Eq. \ref{zfac}).
The T-matrix  with  quasi-particle propagators
 takes the following renormalized form
\begin{eqnarray}
\label{tmren} 
<{\bf p}|T
({\bf P},\Omega)|{\bf p}^{'}> & = & <{\bf p}|V^{ren}
({\bf P},\Omega)|{\bf p}^{'}>   \nonumber \\ & &  
+ 
\int\frac{d^3q}{(2 \pi)^3} <{\bf p}|V^{ren}
({\bf P},\Omega)|{\bf q}> Z_{p_1}Z_{p_2}
\frac{(1-f(\omega_{p_1})-f(\omega_{p_2}))}
{\Omega-\omega_{p_1}-\omega_{p_2}+i\epsilon}
 <{\bf q}|T({\bf P},\Omega)
|{\bf p}^{'}> \ ,  
\end{eqnarray}
with the renormalized interaction $V^{ren}$  given by
\begin{eqnarray}
\label{vren} & & 
<{\bf p}|V^{ren}({\bf P},\Omega)|{\bf p}^{'}> = V({\bf p},{\bf p}^{'})
 + 
 \int\frac{d\omega_1}{2\pi}\int\frac{d\omega_2}{2\pi}
\int\frac{d^3q}{(2 \pi)^3} V({\bf p},{\bf q}) 
B(p_1,\omega_1,p_2,\omega_2,\Omega)
 <{\bf q}|V^{ren}({\bf P},\Omega)|{\bf p}^{'}> \ . 
\end{eqnarray}
The  bare interaction is renormalized by contributions from background 
parts of the spectral functions, this involves integration over energies
 far from the quasi-particle pole. The background part of the spectral 
function is of course necessary to recover sum rules for the particle strength.
However, the full treatment of the renormalized interaction is difficult and
in the following we  replace it by the bare interaction  $V^{ren}\simeq V$
in the gap equation.
The only remnant of the dressing of nucleons in medium are the single 
particle energies
\begin{equation}
\label{spe}
\omega_p=\xi_p+Re\Sigma(p,\omega_p)
\end{equation}
and the $Z_p$ factors in the homogeneous term in the T-matrix equation.
It will turn out that these are the dominant modifications responsible for
shifting the critical temperature. In the vicinity of the pole the T-matrix
equation is dominated by the homogeneous term. In this region the 
interaction for quasi-particles is effectively
 renormalized by a factor $Z_{p_1}Z_{p_2}$.

\subsection{Two-particle pole with renormalized interactions}

\label{renint}

According to the Thouless criterion, a pole in 
the T-matrix at the Fermi energy ($\Omega=0$) 
means that superfluid  long range order sets in (Sect. \ref{Thouless}). 
As mentioned earlier, standard mean-field gap equation uses
the following kernel 
\begin{equation}
\label{mfgapkernel}
V(p,k)\frac{\Big(1-2f(\zeta_k)\Big)}{2\zeta_k}
\end{equation}
with mean-field single particle energies $\zeta_k=k^2/2m +\Sigma_{HF}(k)-\mu$.
It leads to a critical temperature $T_c=5$MeV. One could take
single particle energies $\omega_p$ beyond Hartree-Fock, e.g. from 
BHF or T-matrix calculations. However, the difference between the 
resulting effective masses is of the order 
$Z_p$. Hence we stay at the order of Hartree-Fock effective mass
in the standard gap equation, including the modified single particle energies
(\ref{spe}) only together with $Z_p$ factor renormalization.

The second estimate for the critical temperature is obtained from the
 condition of appearance of the pole in the T-matrix with full 
self-consistent spectral functions. It is equivalent to the following kernel 
in the gap equation
\begin{equation}
\int \frac{d \omega}{2 \pi}
\int \frac{d \omega^{'}}{2 \pi} 
 V({\bf p},{\bf k}) 
\frac{A({ k},\omega-\omega^{'})
A({ k}, \omega)\big(1-f(\omega-\omega^{'})-
f(\omega^{'})\big)}{\omega} \ ,
\end{equation}
where $A(p,\omega)$ is obtained from full self-consistent calculation of 
normal nuclear matter at finite temperature with off-shell propagators 
in the T-matrix ladder \cite{ja}.
It gives a very different value for $T_c=1.6$MeV at $\rho=.45\rho_0$.

Finally we can use the renormalized interaction strength in the homogeneous 
term of the T-matrix equation (Eq. \ref{tmren}). It is equivalent to using
a
renormalized interaction in the kernel of  the quasi-particle 
gap equation
\begin{equation}
\label{poprint}
V(p,k)Z_p^2\frac{\Big(1-2f(\omega_k)\Big)}{2\omega_k} \ ,
\end{equation}
with the single particle energies $\omega_p$ and $Z_p$ factors
obtained from the full self-consistent 
T-matrix calculation (Eq. \ref{spe}). It gives a value of $T_c=2.2$MeV,
much closer to the result of the calculation with full spectral functions.
The $Z$ factor obtained from the full self-energy is not close to $1$, 
at the Fermi energy we find in our model calculation \cite{ja2} 
$Z_{p_f}\simeq0.7$. Such a small value of the renormalization factor
 can explain
the large difference in critical temperatures found
in the standard quasi-particle gap equation and 
in  the one with full spectral function.

\section{Gap equation with dressed propagators}

\label{supersec}
\subsection{Gap equation with full spectral function}

Below the critical temperature a nonzero solution of the gap equation is 
possible.
The kernel of the gap equation is very similar to the  two-particle propagator
in  the T-matrix resummation
  (Eq. \ref{gapfull}). However one of the 
propagators includes the normal self-energy as well as the  off-diagonal
one $\Delta(p)$ \cite{ja2}. 
The full retarded propagator can be expressed using the
normal propagator $G(p,\omega)$ (which includes only the normal self-energy)
\begin{equation}
G_s(p,\omega)=\frac{1}{G(p,\omega)^{-1}+  |\Delta|^2(p) G^{\star}(p,-\omega)}
\ .
\end{equation}
As a result in the kernel of the gap equation 
\begin{eqnarray}
\label{gapfull2} & & 
 \Delta 
({\bf p}) =   - 
 \int\frac{d^3k}{(2 \pi)^3} \int \frac{d \omega}{2 \pi}
\int \frac{d \omega^{'}}{2 \pi} 
 V({\bf p},{\bf k}) 
\frac{A({ k},\omega-\omega^{'})
A_s({ k}, \omega)\big(1-f(\omega-\omega^{'})-
f(\omega^{'})\big)}{\omega} 
 \Delta({\bf k}) \ .  
\end{eqnarray}
 two spectral functions
$A(p,\omega)=-2Im G(p,\omega)$ and $A_s(p,\omega)=-2 Im G_s(p,\omega)$ appear.
Both 
spectral functions can be calculated knowing the two self-energies
$\Sigma(p,\omega)$ and $\Delta(p)$. The off-diagonal self-energy, ie. 
the superfluid gap can be obtained from the gap equation (\ref{gapfull2}).
On the other hand the normal self-energy cannot be calculated in the usual
 T-matrix approximation as above $T_c$. 
This is related to the appearance of 
the 
Cooper instability in the two-particle propagator in the ladder approximation.

\begin{figure}
\centering
\includegraphics[width=0.6\textwidth]{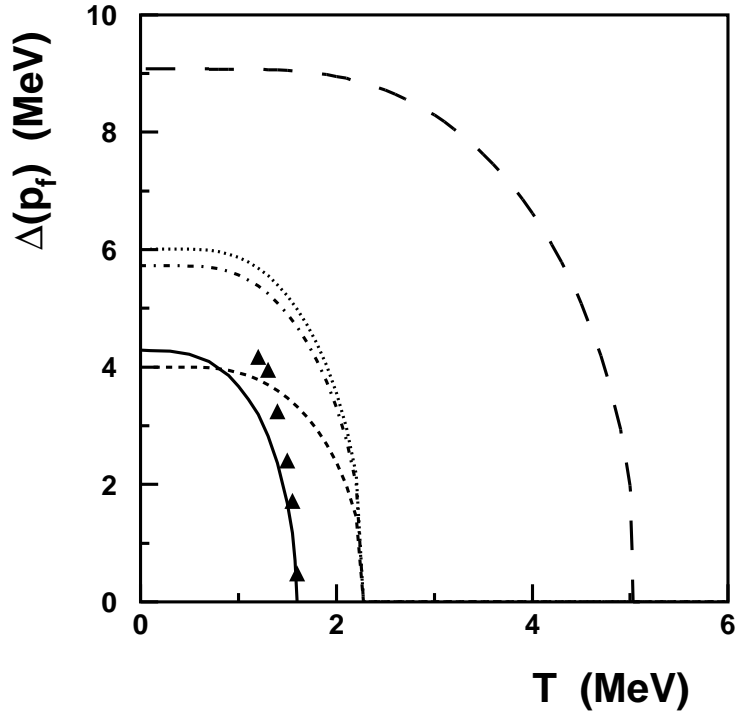}
\caption{Superfluid gap at the Fermi momentum as function of the 
temperature at $\rho=.45\rho_0$. Points denote results obtained from 
self-consistent solution of the ladder self-energy and gap equations. 
The solid line represents results obtained using the imaginary part of 
the self-energy fixed at  $T=1.63$MeV and solving self-consistently 
the Hartree-Fock and the full gap equation (\ref{gapfull2}). The long dashed 
line is the mean-field quasi-particle solution of the gap equation with 
Hartree-Fock single-particle energies (\ref{gapzwyk}). 
The dashed line is the result of the 
quasi-particle approximation (\ref{qpgap2})
 without renormalizing the superfluid
 gap (\ref{naiveE}).
 The dashed-dotted and dotted lines are the results of the quasi-particle 
approximation (\ref{qpgap2})
 with renormalization of the superfluid gap by $Z_p$ (\ref{tezdobre}) and by
$1/(1-O(p,\omega_p))$ (\ref{lepsze}) respectively.}
\label{szcz}
\end{figure}

The T-matrix equation can be modified by introducing also
anomalous propagators in the ladder \cite{hau2}. In Ref. \cite{ja2}
we used a simpler modification of the ladder diagrams, using a mixed 
ladder with
one full propagator and one with only normal self-energy included. 
This means that the singularity of the T-matrix which appears at $T_c$ at
the Fermi energy $\Omega=0$ and zero total momentum of the pair, 
stays there also below $T_c$. This reflects the presence of long range order
\cite{km,hau2}. In such a way we were able to obtain self-consistent 
solutions for the normal self-energy in the ladder approximation and 
for the 
superfluid gap in the mean-field approximation, using at all stages full 
spectral functions $A(p,\omega)$ and $A_s(p,\omega)$, without 
quasi-particle approximation. The results for
the superfluid gap at several temperatures below $T_c$ are indicated by 
triangles in Fig. \ref{szcz}. This calculation cannot be extended 
straightforwardly to zero temperature because at the Fermi energy the
 imaginary part of the self-energy vanishes and the discretization 
of the spectral functions is not possible close to the Fermi energy. 
At the Fermi momentum quasi-particles appear with small width.

\begin{figure}
\centering
\includegraphics[width=0.6\textwidth]{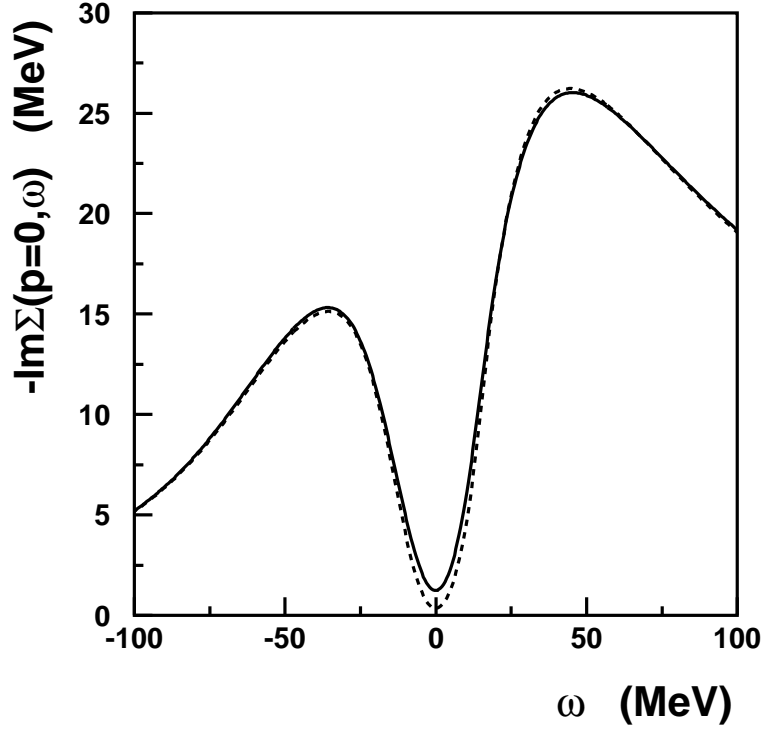}
\caption{Imaginary part of the self-energy at zero momentum for two
temperatures $T=1.7$ (solid line)and $1.2$ MeV (dashed line).}
\label{gpor}
\end{figure}

By comparing the imaginary part of the self-energy at two temperatures 
$T=1.7$MeV and $1.2$MeV one notices that the only modification is  the 
reduction of single-particle width close to the Fermi energy with temperature, 
in agreement with
general properties of Fermi liquids.
Thus we can take as an approximation a temperature independent imaginary 
part of the self-energy. We proceed by taking the imaginary part of the 
self-energy $Im \Sigma(p,\omega)$ 
as calculated at $T=1.63$MeV, slightly above $T_c$. 
Obviously the dispersive contribution 
to the real part of the self-energy
is also fixed (\ref{disper}). The Hartree-Fock energy is obtained from
\begin{equation}
\Sigma_{HF}(p)=\int \frac{d^3k}{(2\pi)^3}V(|{\bf p-k}|/2,|{\bf p-k}|/2)
n(p) \,
\end{equation}
where the momentum distribution is given by
\begin{equation}
n(p)=\int \frac{d\omega}{2\pi} A_s(p,\omega) f(\omega) \ .
\end{equation}
The shift of the Fermi energy from the value $\omega=0$ at $T=1.63$MeV
to keep the density
\begin{equation}
\rho=.45\rho_0=\int \frac{d^3 p}{(2\pi)^3} n(p)
\end{equation}
constant when decreasing the temperature to zero is only $0.2$MeV and has 
negligible effect on the Hartree-Fock energy. Thus the main change in the 
self-energy when decreasing the temperature occurs in the off-diagonal part 
$\Delta(p)$.
At any given temperature the superfluid gap is obtained from the
gap equation with full spectral function (\ref{gapfull2}). 
Where the dependence
 of the kernel of the gap equation on $\Delta(p)$
 enters through the spectral function 
\begin{eqnarray}
\label{As}
A_s(p,\omega)&=&
-2\Bigg(\big(\omega+\xi_p+{\rm Re}\Sigma^{+}(p,-\omega)\big)^2
{\rm Im}\Sigma^{+}(p,\omega) 
+{\rm Im}\Sigma^{+}(p,-\omega) \Delta^2(p) 
+ \big( {\rm Im}\Sigma^{+}(p,-\omega)\big)^2 {\rm Im} \Sigma(p,\omega)\Bigg)/
\nonumber \\& &
\Bigg(\Big(\big(\omega-\xi_p-{\rm Re}\Sigma^{+}(p,\omega)\big)
\big(\omega + \xi_p +{\rm Re}\Sigma^{+}(p,-\omega)\big) 
-
{\rm Im}\Sigma^{+}(p,\omega){\rm Im}\Sigma^{+}(p,-\omega)-\Delta^2(p)\Big)^2
\nonumber \\ & &
+\Big( {\rm Im}\Sigma^{+}(p,\omega)\big(\omega+\xi_p+{\rm Re}\Sigma^{+}(p,
-\omega)\big)
+ {\rm Im}\Sigma^{+}(p,-\omega)
\big(\omega-\xi_p-{\rm Re}
\Sigma^{+}(p,\omega)\big)\Big)^2 \Bigg) \ .
\end{eqnarray}
In the above equation $Im \Sigma$ is fixed and temperature independent, 
according to our approximation, and $Re \Sigma $ depends only very weakly
 on the temperature and the superfluid gap through the Hartree-Fock energy.
The solution of the gap equation using a fixed single particle width 
 down to zero temperature is represented by the solid line in Fig. \ref{szcz}.
As expected close to the temperature where the imaginary part of the
 self-energy was fixed ($T=1.63$MeV) the result is close to the fully 
self-consistent solution denoted by the triangles on the figure. As the
temperature is lowered some deviations appear. It is due to the decrease of the
single particle width close to the Fermi energy in the fully
self-consistent solution (Fig. \ref{gpor}).
 In the following we will compare the 
solution of the gap equation with full spectral function 
to the corresponding quasi-particle approximation.  Since the  
imaginary part of the self-energy is fixed at $T=1.63$MeV,
the properties of the quasi-particle pole ($\omega_p$ and $Z_p$) 
are taken from the self-consistent 
solution at the same temperature.
Our aim is to explain the big difference between the mean-field solution
 with
on-shell propagators (long-dashed line in Fig. \ref{szcz}) 
and the 
gap equation with full spectral function below $T_c$ (solid line).
 For this purpose it is sufficient
to compare the approximation
 with nontrivial but fixed spectral properties with its quasi-particle limit.

The mean-field BCS gap equation is
\begin{eqnarray}
\label{gapzwyk}
 \Delta
({\bf p}) + 
 \int\frac{d^3k}{(2 \pi)^3} \
 V({\bf p},{\bf k}) 
\frac{\Big(1-2f(E_k)\Big)}{2E_k} 
 \Delta({\bf k}) = 0 \ ,
\end{eqnarray}
where $E_k=\sqrt{\zeta_k^2+\Delta(k)^2}$. It gives 
significantly larger values 
for the superfluid gap (long-dashed line in Fig. \ref{szcz})
 than the 
full solution (solid line) in all ranges of temperatures below $T_c$.

\subsection{Quasi-particle approximation for the superfluid spectral function}

\begin{figure}
\centering
\includegraphics[width=0.6\textwidth]{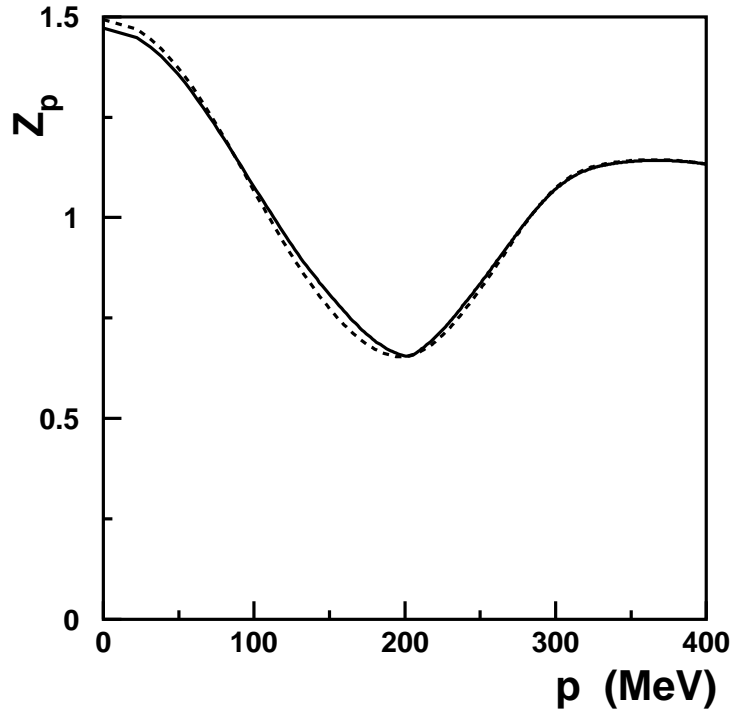}
\caption{Quasi-particle pole renormalization factor for the full spectral 
function $Z^{'}_p$ (Eq. \ref{zsf}) as function of momentum (solid line)
compared to the quasi-particle renormalization in the normal spectral function
$Z_p$ (Eq. \ref{zfac}) (dashed line) for $\Delta(p_F)=2$MeV.}
\label{zfacfig}
\end{figure}

Above $T_c$ we have used the standard form of the quasi-particle 
approximation for the spectral function $A(p,\omega)=2 \pi Z_p
 \delta(\omega-\omega_p)$. Using it we have found a significant reduction 
of the critical temperature with respect to the mean-field approximation. 
To obtain a quasi-particle approximation for the kernel of the gap equation 
(\ref{gapfull2}) with full spectral function we have to construct an 
approximation for the superfluid spectral function $A_s$.
We can write the spectral function (\ref{As})  putting
an infinitesimally small  imaginary part of the self-energy
($Im \Sigma(p,\omega)
=-i\eta$)
\begin{eqnarray}
& &A_s(p,\omega)= 
-2Im\Bigg( 
 \Big( \omega+\xi_p+Re\Sigma(p,-\omega)+i\eta\Big)/   
\Big((\omega-\xi_p-Re\Sigma(p,\omega)) \nonumber \\ & &
(\omega+\xi_p+Re\Sigma(p,-\omega))
  +\Delta^2(p)+i\eta
(2\omega-Re\Sigma(p,\omega)+Re\Sigma(p,-\omega))\Big) \Bigg) \ , 
\end{eqnarray}
with $\xi_p=p^2/2m-\mu$. It is useful to define the even 
$S(p,\omega)=Re\Sigma(p,\omega)+Re\Sigma(p,-\omega)$
and odd $\omega O(p,\omega)=Re\Sigma(p,\omega)-Re\Sigma(p,-\omega)$ 
parts of the real part of the 
 self-energy with respect to the Fermi energy \cite{mahan}.
Notice that we can take $S(p,0)=0$ redefining $\mu$. We have
\begin{eqnarray}
& & A_s(p,\omega)=  
-2Im\Bigg( 
\frac{\omega(1-O(p,\omega))+\xi_p+S(p,\omega)}
{\big( (1-O(p,\omega))-E(p,\omega)\big)
\big(\omega(1-O(p,\omega))+E(p,\omega)\big)+i\eta\omega(1-O(p,\omega))}
 \Bigg) \ ,  
\end{eqnarray}
where
$$
E(p,\omega)=\sqrt{(\xi_p+S(p,\omega))^2+\Delta^2(p)} \ .
$$
The above expression has poles at
$$
\pm\omega=\epsilon_p=\frac{E(p,\epsilon_p)}{1-O(p,\epsilon_p)}=
\frac{E_p}{1-O_p} \ ,
$$
which gives two
quasi-particle contributions to the spectral function on both sides of 
the Fermi energy
\begin{equation}
A_s(p,\omega)=2\pi Z^{'}_p\Bigg(\frac{E_p+\xi_p+S_p}{2E_p}
\delta(\omega-\epsilon_p)+\frac{E_p-\xi_p-S_p}{2E_p}
\delta(\omega-\epsilon_p)\Bigg) \ ,
\end{equation}
with $S_p=S(p,\epsilon_p)$ and
\begin{equation}
\label{zsf}
Z_p^{' \ \ -1}=\Bigg(1
-\frac{\partial \bigg(E(p,\omega)/(1-O(p,\omega))\bigg)}{\partial
\omega}|_{\omega=\epsilon_p}\Bigg)(1-O_p) \ .
\end{equation}
It will be useful to relate this new pole renormalization strength 
to the usual renormalization factor  $Z_p$ (Eq. \ref{zfac}).
We can write
\begin{equation}
\label{zprzy}
Z_p^{' \ \  -1}
= \big(1-\frac{\partial \ \Sigma(p,\omega)}{\partial \omega}|_{\omega=
{\rm sign}(p-p_f)\epsilon_p}\big)+\frac{E_p- {\rm sign}(p-p_f)(\xi_p+S_p)}{E_p}
\frac{\partial S(p,\omega)}{\partial \omega}|_{\omega=\epsilon_p} \ ,
\end{equation}
with ${\rm sign}(x)=\Theta(x)-\Theta(-x)$~.
For small values of the superfluid gap the superfluid quasiparticle position
${\rm sign}(p-p_f)\epsilon_p$ is very close to the position of the 
pole in the normal propagator $\omega_p=\xi_p+Re\Sigma(p,\omega_p)$.
Substituting $\omega_p$ for the energy argument in the first term in Eq.
 (\ref{zprzy}) we get
\begin{equation}
Z_p^{' \ \  -1}- Z_p^{-1}=
\frac{E_p-{\rm sign}(p-p_f)(\xi_p+S_p)}{E_p}
\frac{\partial S(p,\omega)}{\partial \omega}|_{\omega=\epsilon_p} \ .
\end{equation}
The first factor on the right hand side of  the above equation
  is small except close to the Fermi 
energy 
 where the second factor precisely vanishes.
Thus we expect that the new quasi-particle pole renormalization factor 
$Z^{'}_p$ can be approximated by the renormalization factor of the normal
 spectral function $Z_p$ corresponding to the same momentum. A
numerical calculation confirms this to a very good accuracy 
(Fig. \ref{zfacfig}).

The spectral function takes the form
\begin{equation}
A_s(p,\omega)=2\pi Z_p\Bigg(u_p^2
\delta(\omega-\epsilon_p)+v_p^2
\delta(\omega-\epsilon_p)\Bigg) \ ,
\end{equation}
with  coherence factors 
\begin{equation}
u_p^2 (v_p^2)=\frac{\epsilon_p \ +(-) \ (\xi_p+S_p)/(1-O_p)}{2\epsilon_p} \ .
\end{equation}
These coherence factors can be very well approximated by an expression 
similar to the one used in the  usual mean-field gap equation
\begin{equation}
u_p^2 (v_p^2)=\frac{\epsilon_p \ +(-) \omega_p}{2\epsilon_p} \ .
\end{equation}
The approximation works to within $1\%$ for $\Delta(p_f) < 10$MeV.

Substituting the quasi-particle expressions for the spectral functions $A$ 
and $A_s$ into the kernel of the gap equation one obtains
\begin{equation}
\label{qpgap2}
\Delta(p)  =  -\int \frac{d^3 k}{(2 \pi)^3} Z_k^2 V(p,k)\frac{(1-
2f(\epsilon_p))}{2\epsilon_p} \ .
\end{equation}
We have obtained an expression for the gap equation with a kernel very 
similar to the mean-field BCS one (Eq. \ref{mfgapkernel}), but with the 
interaction renormalized by $Z_p^2$ and a different quasi-particle energy
$\epsilon_p$.

In the limit of vanishing gap $\Delta(p)\longrightarrow 0$  all
quasi-particle approximations for the gap equation reduce to the Thouless
condition for the
renormalized T-matrix equation (Eq. \ref{poprint}).
 Accordingly we have the same condition for $T_c$
as discussed in Sect. \ref{renint}.

\subsection{Quasi-particle energies}

In order to relate our quasi-particle gap equation (\ref{qpgap2}) 
to usual approaches
 we have to obtain an approximation for the quasi-particle energies
\begin{equation}
\label{epsiloneq}
\epsilon_p=\frac{\sqrt{(\xi_p+S_p)^2+\Delta^2(p)}}{1-O_p} \ .
\end{equation}
To extract the positions
  $\pm\epsilon_p$
 of the poles 
of the spectral 
function one has to know the real part of the self energy for energies on
 both sides of the Fermi energy.
We would like to obtain an expression which could be used as a correction to 
 calculations using quasi-particle approximation like BHF.

In BHF
approaches one calculates a single particle energy which includes
dispersive corrections to the position of the quasi-particle pole 
(Eq. \ref{spe}) in the usual spectral function $A(p,\omega)$.
A simple approximation would be to use $\omega_p$ instead of $\zeta_p$
in the expression for the quasi-particle energies in the superfluid 
\begin{equation}
\label{naiveE}
\epsilon_p=\sqrt{\omega_p^2+\Delta^2(p)} \ .
\end{equation}
\begin{figure}
\centering
\includegraphics[width=0.6\textwidth]{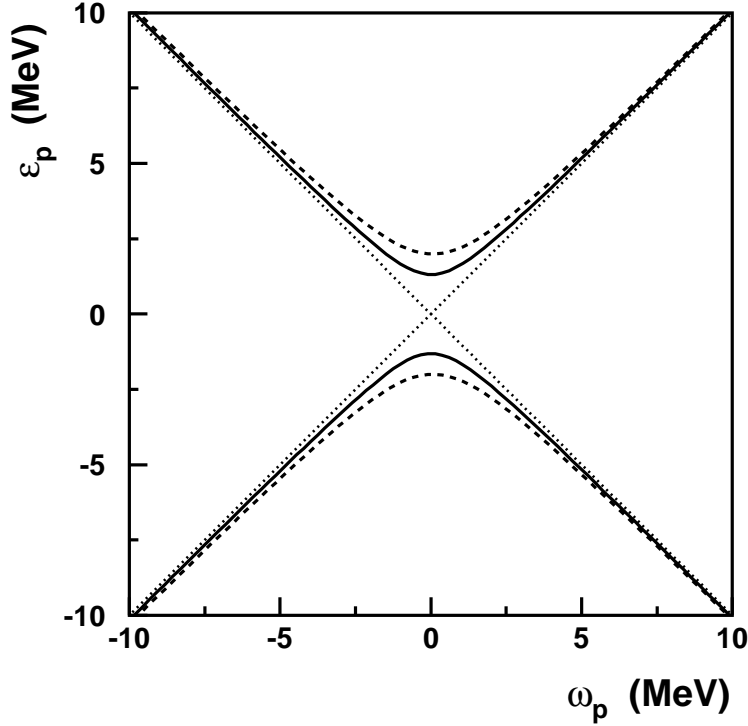}
\caption{Position of the superfluid 
quasi-particle pole $\epsilon_p$  (solid line) 
as function of quasi-particle energy compared to the naive 
expression $\sqrt{\omega_p^2
+\Delta^2(p)}$ (dashed line) (Eq. \ref{naiveE}) for $\Delta(p_f)=3$MeV.
 The quasi-particle energies obtained using 
renormalized superfluid gap 
$\sqrt{\omega_p^2+\Delta(p)^2/(1-O_p)^2}$ (Eq. \ref{renoD})
cannot be distinguished from the solid 
line on the scale of the figure.}
\label{enerfig}
\end{figure}
Clearly the above expression does not give the right value for the
energy gap at the Fermi momentum (Fig. \ref{enerfig}).
The energy gap at $p_F$ should be
\begin{equation}
\epsilon_{p_F} =\frac{\Delta(p_F)}{1-O_{p_F}} 
  = Z_{p_F}\Delta(p_F) \ .
\end{equation}
Let us define a renormalized energy gap
\begin{equation}
\label{renoD}
\tilde{\Delta}(p)=\frac{\Delta(p)}{1-O_p} \ .
\end{equation}
Results for the quasi-particle poles using this renormalization of the 
superfluid order parameter
\begin{equation}
\label{lepsze}
\epsilon_p=\sqrt{\omega_p^2+\tilde{\Delta}^2(p)} 
\end{equation}
are indistinguishable from the solution (\ref{epsiloneq}) 
in Fig. \ref{enerfig}.
It should be noted that for the calculation of the factor $(1-O_p)$ we 
can take $O_p=O(p,\epsilon_p)\simeq O(p,\omega_p)$. So that the 
approximation for $\epsilon_p$
can be expressed using the quasi-particle pole energy  $\omega_p$ and the 
values of the real part of the self energy at $\pm \omega_p$.
However, usually  in BHF calculations one does not calculate the factor 
$1-O_p$, since  the value and the derivative of the real part of the
self-energy is known only at $\omega_p$ and  not at $-\omega_p$. 
In this case the position of the poles in the superfluid can be approximated by
\begin{equation}
\label{tezdobre}
\epsilon_p=\sqrt{\omega_p^2+\hat{\Delta}^2(p)} 
\end{equation}
with the renormalization of the superfluid gap  taken 
as
\begin{equation}
\label{ZrenoD}
\hat{\Delta}(p)={\Delta(p)}Z_p \ .
\end{equation}
$\hat{\Delta}(p)$ is very close to $\tilde{\Delta}(p)$ around the Fermi
 energy. Some differences appear only around $p=0$ 
(for the assumed interaction), but this region is less important for the
solution of the gap equation.
Again the energies obtained using expression (\ref{tezdobre})
are indistinguishable from the solid line in Fig. \ref{enerfig}.

\subsection{Results for superfluid gap in quasi-particle approximation}

In  previous sections we  derived a quasi-particle approximation for 
the kernel of the superfluid gap equation (Eq. \ref{gapfull2}).
The results of the full gap equation with integration of over the energies 
in the fermion propagators (solid line in Fig. \ref{szcz}) can be compared to 
 quasi-particle results.
The dotted line in  Fig. \ref{szcz} 
represents the solution of the gap equation 
(\ref{qpgap2}) with  renormalization of the energy gap $\tilde{\Delta}(p)$
for the calculation of the superfluid quasi-particle energies 
(Eq. \ref{lepsze}).
The dashed-dotted line represents an analogous calculation but with the 
energy gap $\hat{\Delta}(p)$ (Eq. \ref{ZrenoD}).
Both quasi-particle approximation are very close to each other 
with critical temperature $T_c=2.2$MeV. The value of $T_c$ is
 the same as obtained in Sect. \ref{renint} 
using renormalized interactions 
in the ladder approximation. This value of critical temperature
is much closer to the one
 obtained from the self-consistent solution $T_c=1.6$MeV
than was the mean-field value
$T_c=5$MeV. Also the value of the superfluid gap is much closer to the
one calculated using full of-shell spectral functions (Table \ref{table1}).
The  superfluid gap at zero temperature in the quasi-particle approximation
$\Delta(p_F)=6$MeV is much closer to the solution of the off-shell gap 
equation $\Delta(p_F)=4.3$MeV than the one obtained from the mean-field BCS
solution $\Delta(p_F)=10.8$MeV. From the results in Table \ref{table1}
 we can notice 
that the ratio between   the superfluid gap at
 zero temperature and the critical temperature
 is  larger in the solution of the  full gap equation 
(\ref{gapfull2}) and in 
quasi-particle approximations (\ref{qpgap2})
with renormalization of the energy gap (\ref{lepsze} or \ref{tezdobre}), than
in the mean-field solution (\ref{gapzwyk})
and in quasi-particle approximations without
renormalization of the superfluid energy gap (\ref{qpgap2}, \ref{naiveE}).
In the weak coupling limit we have \cite{mahan,fw}
\begin{equation}
\frac{\Delta(p_F)|_{T=0}}{T_c}=\pi e^{-\gamma}\simeq1.76 \ .
\end{equation}
\begin{center}
\begin{table}
\caption{Values for the critical temperature and superfluid gap at zero 
temperature for different approximations.}
\label{table1}
\begin{tabular}{l|ccc}
\hline
approximation & ~$T_c$ (MeV)~~ & ~~$\Delta(p_F)_{T=0}$ (MeV)~~ &
 ~~$\Delta(p_F)/T_c$~~ \\
\hline \hline
off-shell gap &  1.6 & 4.3 & 2.7 \\
equation (\ref{gapfull2}) & & & \\ \hline 
quasi-particle app. with & 2.2 & 6.0 & 2.7 \\
$\tilde{\Delta}(p)$ (Eqs. \ref{qpgap2}, \ref{lepsze}) & & & \\ \hline
quasi-particle app. with & 2.2 & 5.7 & 2.6 \\
$\hat{\Delta}(p)$ (Eqs. \ref{qpgap2}, \ref{tezdobre}) & & & \\ \hline
quasi-particle app. with & 2.2 & 4.0 & 1.8 \\
${\Delta}(p)$  (Eqs. \ref{qpgap2}, \ref{naiveE}) & & & \\ \hline
mean-field BCS & 5.0 & 10.8 & 2.2 \\
(Eq. \ref{gapzwyk}) & & & \\ \hline
weak coupling BCS & - & - & 1.76 \\ \hline
weak coupling BCS with & - & - & 2.5 \\
 renormalized energy gap &    & & \\  \hline
\end{tabular}\\
\end{table}
\end{center}
If the energy gap $\tilde{\Delta}(p_F)$ is renormalized with respect to the 
superfluid gap $\Delta(p_F)$ (the off diagonal self-energy) we have
\begin{equation}
\frac{{\Delta}(p_F)|_{T=0}}{T_c}
=\frac{\tilde{\Delta}(p_F)|_{T=0}(1-O_{p_F})}{T_c}
=\frac{\tilde{\Delta}(p_F)|_{T=0}}{T_c Z_{p_F}}
\simeq\frac{1.76}{Z_{p_F}}\simeq 2.5\ .
\end{equation}
Ratios of the superfluid gap and of the critical temperature 
for the solution of the full gap equation and quasi-particle 
approximations with renormalized superfluid gap is close to $1.76/Z_{p_F}$
(Table \ref{table1}).
 On the other hand mean-field gap equation and quasi-particle 
approximation without renormalization of $\Delta(p)$ give $\Delta(p_F)|_{T=0}
/T_c$ closer to $1.76$.
These relation can be fulfilled only approximately since
 we are far from the region of 
applicability of the weak coupling BCS solution.


\section{Conclusions}
\label{conc}

Numerical solution of the gap equation with full spectral functions
(\ref{gapfull2}) shows a strong reduction of  the superfluid gap and of
the critical temperature with respect to the mean-field BCS solution
(\ref{gapzwyk}). To understand this effect we constructed a quasi-particle 
approximation for the full gap equation.
The effects of nontrivial spectral functions can be approximated using
a renormalized strength of the interaction $V(p,k)Z_k^2$.
Also the energy gap in the calculation of the quasi-particle poles in the
superfluid is renormalized. For the  renormalization of the gap we used two
expression $\Delta(p)/(1-O_p)$ and $\Delta(p) Z_p$. In both cases we
obtained results for $\Delta(p)$ and $T_c$ much closer to the full solution 
than the mean-field approximation. Having at one's disposal only the single
particle energies $\omega_p$ and the $Z_p$ factors obtained from a realistic
BHF type calculation, the  gap equation can be corrected.
A reduced interaction strength $V(p,k) Z_k^2$ must be taken 
 and a factor $Z_p$ appears between the 
energy gap $\hat{\Delta}(p)$  and the 
off-diagonal self-energy $\Delta(p)$. This is equivalent to solving a gap 
equation for $\hat{\Delta}(p)$ with reduced interaction $V(p,k) Z_p Z_k$.
In the weak coupling limit $\hat{\Delta}(p)|_{T=0}\simeq 1.76 T_c$ but
${\Delta}(p)|_{T=0}\simeq 1.76 T_c/Z_{p_F}$ .

In the illustrative model here presented the scattering corrections are  
very strong ($Z_{p_F}\simeq .7$) and there are still some differences between 
the improved quasi-particle approximation and the full solution of the gap
 equation. We expect that at normal nuclear density where the $Z$ factor is
 closer to $1$, quasi-particle approximation with renormalized interaction 
strength would be much closer to the full solution. This is the case 
for neutron matter where
 effects of renormalization of quasi-particle poles and 
shifts in single-particle energies are smaller
 \cite{baldop}. In future work we plan to 
investigate the effects of the renormalization of the interaction by 
background corrections (\ref{vren}).     

 In this investigation we used 
the bare potential for the two-particle irreducible vertex. 
However, it is known that 
polarization effects reduce  the superfluid gap by a factor $\simeq 3$ 
\cite{wambach}. We must conclude that although there are no ladder corrections
to the interaction in the gap equation, other many-body effects modify the
effective interaction. Due to the exponential dependence of the
gap solution on  the strength of the interaction, these usually neglected 
corrections modify strongly superfluid parameters in nuclear matter.

\acknowledgements

\vspace{2cm}
This work was partly supported by the National Science Foundation
under Grant PHY-9605207.

\appendix




\begin{references}
\bibitem{migdal} A.B. Migdal, {\it Theory of Finite Fermi Systems}
(John Wiley and Sons, New York, 1967).
\bibitem{wambach} J.W. Clark, C.G. K\"allman, C.H. Yang and
D.A. Chakkalakal, Phys. Lett.  {\bf B61} 331 (1976);
T.L. Ainsworth, J. Wambach and D. Pines, Phys. Lett.
{\bf B222} 173 (1989); J. Wambach, T.L. Ainsworth and D. Pines, Nucl. Phys.
{\bf A555} 128 (1993);  
 H.-J. Schulze, J. Cugnon, A. Lejeune, M. Baldo and
U. Lombardo, Phys. Lett. {\bf B375} 1 (1996).
\bibitem{ja2} P. Bo\.zek, Nucl. Phys. {\bf A657} 187 (1999).
\bibitem{bhf} M. Baldo, I. Bombaci, L.S. Ferreira, G. Giansiracusa and
U. Lombardo, Phys. Rev. C {\bf 43} 2605 (1990).
\bibitem{vo} B.E. Vonderfecht, W.H. Dickhoff, A. Polls and A. Ramos,
Nucl. Phys. {\bf A555} 1 (1993). 
\bibitem{roepke} T. Alm, G. R\"opke, A. Schnell, N.H. Kwong and
S. K\"ohler, Phys. Rev. C{\bf 53} 2181 (1996);
 A. Schnell, T. Alm and G. R\"opke, Phys. Lett. {\bf B387} 443
 (1996).
\bibitem{d1}  W.H. Dickhoff, {\em Phys. Rev.} C{\bf 58} 2807 (1998).
\bibitem{ja}
P. Bo\.zek, Phys. Rev. C{\bf 59} 2619 (1999).
\bibitem{thouless} D.J. Thouless, Ann. Phys. (N.Y.) {\bf 10} 553 (1960).
\bibitem{km} L.P. Kadanoff and P.C. Martin, Phys. Rev. {\bf 124} 670 (1961).
\bibitem{hau2} R. Hausmann, Zeit. F\"ur Phys. {\bf B91} 291 (1993);
 R. Hausmann, Phys. Rev. {\bf B49} 12975 (1994). 
\bibitem{pn} P. Nozi\`ere, {\it Theory of Interacting Fermi Systems},
(New York, Benjamin, 1994)
\bibitem{abrikosov} A.A. Abrikosov, L.P. Gorkov and I.E. Dzyaloshinski, 
{\it Methods of Quantum Field Theory in Statistical Physics}
(Prentice-Hall, Englewood Cliffs, 1963).
\bibitem{mahan}J.R. Schrieffer, {\it Theory of Superconductivity}
(W.A. Benjamin, Inc., Massachusetts, 1964);
G.D. Mahan, {\it Many-Particle Physics} 
(Plenum Press, New York, 1981).
\bibitem{fw} A.L. Fetter, J.D. Walecka, {\it Quantum Theory of Many-Particle 
Systems}, (McGraw-Hill, New York, 1971)
\bibitem{baldop} M. Baldo and A. Grasso, nucl-th/003039.
\end{references}
\end{document}